\documentclass{article}
\usepackage{epsfig}
\begin{document}
\baselineskip=16pt
\begin{center}
{\large \textbf{THE EQUIVALENCE BETWEEN DIFFERENT DARK (MATTER) ENERGY SCENARIOS}}\\
\vspace{2cm}
ARBAB I. ARBAB \\
\vspace{1cm} {\small  Department of Physics, Teachers'
College, Riyadh 11491, P.O.Box 4341, Kingdom of Saudi Arabia \\
Comboni College for Computer Science, P.O. Box 114, Khartoum, Sudan}\\
E-mail: arbab@ictp.trieste.it\\
\end{center}
\vspace{1cm} {\small \textbf{Abstract}. We have shown that the
phenomenological models with a cosmological constant of the
 type $\Lambda=\beta\left(\frac{\ddot R}{R}\right)$ and $\Lambda=3\alpha H^2$,
 where $R$ is the scale factor of the universe and $H$ is the Hubble
 constant, are equivalent to a quintessence model with a scalar ($\phi$) potential  of the form
  $V\propto \phi^{-n},\ n$ = constant. The equation of state of the
  cosmic fluid is described by these  parameters ($\alpha, \beta,
  n$) only. The equation of state of the cosmic fluid (dark energy) can be determined by any of these parameters.
  The actual amount of dark energy will define the equation of state of the cosmic fluid.
  All of the three forms can give rise to cosmic acceleration depending the amount of dark energy in the universe.}
 \vspace{.5in}
\\
\centerline{\textbf{ 1. Introduction}}
\\
Recent observation of the
Hubble diagram for supernovae Ia indicates that the expansion of
the universe is accelerating at the present epoch (Perlmutter et
al., 1998; Riess et al., 1998). This apparent acceleration is
attributed to a dark energy residing in space itself, which also
balances the kinetic energy of expansion so  as to give the
universe zero spatial curvature, as deduced from the cosmic
microwave background radiation (CMBR). Cosmologists have proposed
several dark energy models to explain the present cosmic
acceleration. This dark energy was important in the past as it is
now. It might have played a part in limiting the formation of
largest gravitationally bound structures. One can model the dark
energy by a cosmological constant (or a vacuum decaying energy).
However, not all vacuum decaying cosmological models predict this
acceleration. One way to account for such an acceleration is to
propose a kind of scalar field known as quintessence dominating
the universe today. Quintessence offers a possible explanation for
the observed acceleration of the universe without resorting to the
cosmological constant. In essence, quintessence endeavors to
replace the \emph{static} cosmological constant with a dynamical
negative pressure component. However, quintessence models exhibit
an event horizon which poses a serious problem for string theory
(Fichler et al., 2001). The quintessence is supposed to obey an
equation of state of the form $ p_Q=\omega_Q\rho_Q\ $ and
$\omega_Q=-1$ corresponds to vacuum energy density.

In this brief letter we show that the three forms of the
cosmological constants, viz., $\Lambda=\beta\left(\frac{\ddot
R}{R}\right)$ and $\Lambda=3\alpha H^2$ , where $\beta, \alpha$
are constants, and the quintessence model with a scalar ($\phi$)
potential  of the form $V\propto \phi^{-n},\ n$ = constant are
equivalent. We thus see that no one is more fundamental than the
others. Analysis shows that $\beta$ and $\alpha$ determine the
equation of state of the corresponding quintessence.
\\
\centerline{\textbf{2. The Field Equations}}
\\
The Einstein field equations with a variable cosmological
constant, and energy conservation law yield
\begin{equation} \left(\frac{\dot R}{R}\right)^2+\frac{k}{R^2}=\frac{8\pi}{3}
G\rho+\frac{\Lambda}{3}
\end{equation}
\begin{equation}
\frac{\ddot R}{R}=-\frac{4\pi G}{3}(\rho+3p)+\frac{\Lambda}{3}
\end{equation} and
\begin{equation}
\dot\rho+3\frac{\dot R}{R}( p+\rho)=-\frac{\dot\Lambda}{8\pi G}.
\end{equation}
Using the equation of the state
\begin{equation}
p=(\gamma-1)\rho, \qquad \ 1\le \gamma \le 2 \ ,\end{equation}
eq.(2) can be written as
\begin{equation} \frac{\ddot R}{R}=\frac{8\pi
G}{3}\left(1-\frac{3}{2}\gamma\right)\rho +\frac{\Lambda}{3}.
\end{equation} We
see from eq.(3) that a variable $\Lambda$ induces a term
representing the rate of decay of vacuum energy into
matter/radiation (or the rate of generation of entropy). It is
thus apparent that for a decreasing $\Lambda$ the entropy
increases.
\\
\centerline{\textbf{\textbf{3.} The Observable Cosmological
Parameters}}
\\
 Following Arbab (2003, 1997), we consider following forms of $\Lambda$
\begin{equation}
\Lambda=\beta\left(\frac{\ddot R}{R}\right),
\end{equation}
 and
\begin{equation}
\Lambda=3\alpha H^2,
\end{equation}
where $\alpha$ and $\beta$ are undetermined constants.

For the matter-dominated flat universe ($\gamma=1, k=0$). Thus
eqs.(1), (2), (4) and (5) give (Arbab, 2003)
\begin{equation}
R(t)=C t^{\frac{\beta-2}{\beta-3}}\ \ ,\
 \qquad C=\rm const.\ , \ \beta\ne  3 \end{equation}
 Then eq.(6) becomes
\begin{equation}
\Lambda(t)=\frac{\beta(\beta-2)}{(\beta-3)^2}\frac{1}{t^2}\ \ ,
\qquad
 \beta\ne 3, \end{equation}
and eq.(3) yields
 \begin{equation}
 \rho(t)=\frac{(\beta-2)}{(\beta-3)}\frac{1}{4\pi
Gt^2}\ \ , \qquad \beta\ne 3.
\end{equation}
The deceleration parameter is given by
\begin{equation} q=-\frac{\ddot RR}{\dot
R^2}=\left(\frac{1}{2-\beta}\right)\ , \qquad \ \ \beta\ne 2
\end{equation}
It is clear that for a positive energy density $\beta>3$ so that
$\Lambda >0$ and $q<0$. Hence, we obtain a cosmic acceleration
with a minimal requirement.
\\ The mass density parameter of the universe is given by
\begin{equation} \Omega_m=\frac{2}{3}\frac{(\beta-3)}{(\beta-2)}\
, \qquad \ \beta\ne 2 \end{equation} and the the vacuum density
parameter
\begin{equation} \Omega_v=\frac{\beta}{3(\beta-2)}\ \ , \qquad \
\beta\ne 2, \end{equation} so that $\Omega_m+\Omega_v=1$, as
preferred by inflation. Note that all other cosmological
parameters depend on the constant ($\beta$). The case $\beta=2$
defines a static universe, which is physically unacceptable  for
the present universe. For $\beta > 0 $ eq.(13) implies that
$\Omega_v>\frac{1}{3}$. We see that the universe will be
ultimately driven into a de-Sitter phase of exponential expansion
at the present epoch if $\Omega_v\rightarrow 1$ (or $\beta
\rightarrow 3$) (Arbab, 2003).
\\
\centerline{\textbf{4. The Quintessence, the Cosmological Constant
and Equation of State}}
\\
In general, the most important difference of a dark energy
component to a cosmological constant is that its equation of state
can be different form $p=-\rho$, generally implying a
time-variation. \\
Using eq.(5) one can write eq.(14) as
\begin{equation}
\Lambda=\left(\frac{\beta}{3-\beta}\right)\left(1-\frac{3}{2}\gamma\right)8\pi
G\rho \ ,\qquad \ \beta\ne 3 .
\end{equation}
It is evident that when $\gamma=\frac{2}{3}$ the cosmological
constant vanishes ($\Lambda=0$). Thus if the universe is dominated
by strings today, the cosmological constant must vanish! \\
For the matter-dominated universe ($\gamma=1$) eq.(14)
gives\begin{equation}
\Lambda=\left(\frac{\beta}{\beta-3}\right)4\pi G\rho\ .
\end{equation} It is evident from the above equation that an empty
universe ($\rho=0$) would imply a vanishing cosmological constant ($\Lambda=0$). \\
For the radiation-dominated epoch ($\gamma=\frac{4}{3}$) eq.(14)
gives
\begin{equation} \Lambda=\left(\frac{\beta}{\beta-3}\right)8\pi
G\rho\ . \end{equation} Now eq.(1) and (7) give (for $p=0, k=0$)
\begin{equation}
 \Lambda=\left(\frac{\alpha}{1-\alpha}\right)8\pi G\rho.
\end{equation}
Comparison of eq.(17) with eq.(15) yields
\begin{equation}
\alpha=\frac{\beta}{3(\beta-2)}.
\end{equation}
Hence, eqs.(11) and (18) yield

\begin{equation}
  q=\frac{1}{2}(1-3\alpha).
\end{equation}
Now a cosmic acceleration with a positive cosmological constant
requires
\begin{equation}
\frac{1}{3}<\alpha <1.
\end{equation}
Using eq.(15), eq.(5) reads
\begin{equation}
\frac{\ddot R}{R}=\frac{4\pi
G}{3}\left(\frac{3}{\beta-3}\right)\rho\ ,
\end{equation}
i.e.  if $\beta>3$ then the cosmic acceleration is proportional to
the factor $\left(\frac{1}{\beta-3}\right)\rho$.

Very recently, Majern$\rm\acute{i}$k (2001, 2002) considered a
Friedmann's model with an alternative $\Lambda$-part, as
representing a form of the quintessence. He assumed that $\Lambda$
is proportional to the stress-energy scalar ($T$), viz.
\begin{equation}
\Lambda= 8\pi \kappa  G T,
\end{equation}
where $\kappa$ is a free parameter and $T=\rho$ in the present
epoch.

For  quintessence, the equation of state takes the form
\begin{equation}
p_Q=\omega_Q\rho_Q\ ,\qquad -1<\omega_Q<0\ ,
\end{equation}
so that for eq.(22) one finds (Majern$\rm\acute{i}$k, 2001, 2002),
\begin{equation}
\omega_Q=-\frac{\kappa}{1+\kappa}.
\end{equation}
Unlike the ordinary ideal fluid, the  equation of state of the
quintessence will depend only on  the amount of ordinary matter
and/or dark energy involved. Thus, if one determines $\Omega_v$
then $\omega_Q$ can be calculated.  Comparing eqs.(22) and (15)
one reveals that
\begin{equation}
\kappa=\frac{\beta}{2(\beta-3)} .
\end{equation}
and from eq.(24) one concludes that
\begin{equation}
\omega_Q=-\frac{\beta}{3(\beta-2)}\ \ , \ {\rm or}\qquad
\omega_Q=-\Omega_v,
\end{equation}
using eq.(13). Equation (25) can be inverted to define $\beta$ as
\begin{equation}
\beta=\frac{6\omega_Q}{1+3\omega_Q}.
\end{equation}
Similarly, from eqs.(17), (18), (24) and (25) one finds
\begin{equation}
\alpha=-\omega_Q\ \ ,\ {\rm or} \qquad
\alpha=\frac{\kappa}{1+\kappa}
\end{equation}
Therefore, $\beta$ (or $\alpha$ ) defines the equation of state of
the corresponding equivalent quintessence (`` dark energy ").
 The constrain that $\omega_Q< -0.6$  \ implies\  $\Omega_v > 0.6 $.
 This dictates that $ \beta< 4.5$. Thus one has the stringent constraint on $\beta$,
viz., $3<\beta<4.5$  and a similar one on  $\alpha$ as $0.6
<\alpha <1$.  Consequently, one has $-1<q<-0.4$. As remarked by
Majern$\rm\acute{i}$k that when $\Omega_m\rightarrow 0$ then
$\kappa\rightarrow\infty$ and $\omega_Q\rightarrow -1$; here we
have  as $\beta\rightarrow 3$ (or $\alpha\rightarrow 1)$,
$\Omega_v\rightarrow 1$ (de Sitter universe).\\ We would like to
remark here that $|\omega_Q|$ is nothing but the vacuum energy
parameter, as evident from eq.(7) and the fact that
$\rho_v=\frac{\Lambda}{8\pi G}$. Measuring the dark energy
equation of state will rely very much on the future observations.
Thereafter, one can decide wether quintessence is indeed an
acceptable explanation of the dark energy of the universe.
\\
\centerline{\textbf{5. Cold Dark Matter Model}}
\\
 In cold dark
matter model (CDM) the dark energy interacts only with itself and
gravity. Its equation of state is given by
\begin{equation}
 p_X=\omega_X\rho_X,
\end{equation}
and if $\omega_X$=constant, its energy density is given by
\begin{equation}
\rho_X\propto R^{-3(1+\omega_X)}.
\end{equation}
However, in a scalar field model the parameter $\omega_X$ is
derived from the field  model.\\ Ratra \& Quillen (1992) and Brax
\& Martin (2002) considered a scalar field potential ($V)$ of the
form
\begin{equation}
V(\phi)=\frac{C}{\phi^{n}},
\end{equation}
where $n, \ C$ are some constants. They have found that
\begin{equation}
\omega_X=-\frac{2}{2+n}.
\end{equation}
Comparison of eq.(32) with eq.(26) shows that
\begin{equation}
n=   \frac{ 4(\beta-3)}{\beta}.
\end{equation}
Consequently, an accelerated expansion ($\beta >3)$ requires $n >
0$. This constraint allows the energy density of the scalar field
to roll down very slowly after inflation, but at a still high
red-shift, and it had a relatively small value, so it has not
disturbed the CMBR, but eventually dominates and  the universe
acts as if it had a cosmological constant that varying slowly with
time (and possibly space)(Peebles, 2002). We observe that in the
limit where $n\rightarrow 0$ the scalar field energy density
mimics a cosmological constant.The cosmological consequences of
the models discussed above are the same for the present era.
Therefore, no one of them is more fundamental than the others.
\\
\centerline{\textbf{Acknowledgements}}
\\
I would like to thank Comboni College for providing a research
support for this work.
\\
\centerline{\textbf{References}}
\\
Perlmutter, S.,  et al.: 1998, \emph{Nature} \textbf{391}, 51. \\
Riess, A.G., et al.: 1998,  \emph{Astron. J}. \textbf{116}, 1009.\\
Fichler, W., et al.: 2001, xxx.lanl.gov/abs/hep-th/0104181\\
Majern$\rm \acute{i}k$, V.: 2001,  \emph{Phys. Lett. A}\textbf{282}, 362., xxx.lanl.gov/abs/gr-qc/0201019\\
Arbab, A.I.: 2003, \emph{Class. Quantum Gravit.} \textbf{20}, 93.\\
 Arbab, A.I.: 1997, Gen. Rel. Gravit. \textbf{29}, 61.\\
 Ratra, B., and Quillen, A.: 1992, \emph{MNRAS}. \textbf{259}, 738.\\
Ratra, B., and Peebles, P.J.E.: 1988, \emph{Phys.Rev. D}\textbf{37}, 3406. \\
 Brax, P., and Martin, J.: 2002, xxx.lanl.gov/abs/astro-ph/0210533.\\
 Peebles, P.J.E.: 2002, xxx.lanl.gov/abs/astro-ph/0207347.\\
\label{lastpage}
\end{document}